\documentclass[12pt,a4paper]{article}
\usepackage[]{amsmath,amssymb}
\usepackage{graphics,epsfig}
\begin{document}
\date{}
\title{Entropy inequalities from reflection positivity}
\author{H. Casini\footnote{e-mail: casini@cab.cnea.gov.ar}\\
{\sl Centro At\'omico Bariloche and Instituto Balseiro}\\ 
{\sl 8400-S.C. de Bariloche, R\'{\i}o Negro, Argentina}}
\maketitle
\begin{abstract}
 We investigate the question of whether the entropy and the Renyi entropies of the vacuum state reduced to a region of the space can be represented in terms of correlators in quantum field theory. In this case, the positivity relations for the correlators are mapped into inequalities for the entropies. We write them using a real time version of reflection positivity, which can be generalized to general quantum systems. Using this generalization we can prove an infinite sequence of inequalities which are obeyed by the Renyi entropies of integer index. There is one independent inequality involving any number of  different subsystems. 
 In quantum field theory the inequalities  acquire a simple geometrical form and are  consistent with the integer index Renyi entropies being given by vacuum expectation values of twisting operators in the Euclidean formulation. Several possible generalizations and specific examples are analyzed. 
\end{abstract}

\section{Introduction}
The quantum entropy satisfies several inequalities which have found a variety of applications in different areas of physics \cite{petz}. For example, they play a key role in the recent developments on quantum information theory and quantum computation \cite{qit}, and are important for the statistical mechanics of extended systems such as quantum spin lattices \cite{spin}. Other applications range from the renormalization group irreversibility in quantum field theory \cite{nu}, to black hole information loss \cite{bh}.  

In this paper we find a new series of inequalities for the Renyi entropies of integer index which involve a modular reflection, and which are inspired in ideas related to the entanglement of the vacuum state in a relativistic quantum field theory (QFT). Let $\rho_A$ be  the density matrix  corresponding to the vacuum state reduced to a region $A$ of the space,    $S(A)=-tr(\rho_A \log \rho_A)$ its entanglement entropy, and $S_n(A)=-1/(n-1)\log (\textrm{tr}\rho^n)$ the Renyi entropies, with $S_1(A)=S(A)$.  In recent years some properties of the entanglement in the fundamental state, measured by these quantities, have been connected with several effects in condensed matter physics and QFT.  This subject has now become an active area of research \cite{dt}. 

Our motivation for studying these inequalities starts with an observation on the expression for the  entropies of the free massless fermion field in one spatial dimension  \cite{fermion,ch2} (see also \cite{ccc,glio}). This is up to now the 
only exact example in QFT where the function $S(A)$ is known for any number of connected components in $A$. For one chirality we have,
\begin{equation}
S((a_1,b_1)...(a_p,b_p))=\frac{1}{6 }\left( \sum_{i,j} \log | a_i-b_j |-\sum_{i<j}
\log |a_i -a_j |-\sum_{i<j}\log | b_i -b_j |-p\,\log\epsilon \right)\,,\label{confor}
\end{equation}
where the set is formed by $p$ disjoint intervals $(a_i,b_i)$ on a spatial line, with $a_i<b_i<a_{i+1}$. Here $\epsilon$ is a short distance ultraviolet cutoff.
Somewhat surprisingly, an exponential of this expression can be written in terms of the correlator of the free field itself. Writing $q=\pm 1$ for the chirality of $\Psi$, we have
\begin{eqnarray}
\langle 0 \vert  (-i q) \Psi^\dagger(a_1) \Psi(b_1) ...(-i q) \Psi^\dagger(a_p)\Psi(b_p)  \vert 0 \rangle =\frac{(-1)^p}{(2\pi)^p} \sum_{P} \sigma(P)\frac{1}{a_1-b_{P(1)}}...\frac{1}{a_p-b_{P(p)}}\nonumber \\\hspace{4.5cm}=\frac{1}{(2\pi)^p}\frac{\prod_{i< j} \vert a_i - a_j\vert \prod_{i< j}\vert b_i -b_j \vert}{\prod_{i,j}\vert a_i-b_j \vert}=c^p e^{-6  S((a_1,b_1)...(a_p,b_p))}\,,\label{fermi}
\end{eqnarray} 
where the sum in the first line follows from Wick's theorem and is over the permutations $P$ of $1,2...p$, with $\sigma(P)$ the permutation signature. $\vert 0\rangle$ is the vacuum state and  $c=1/(2\pi\epsilon)$ is a cutoff dependent constant which also gives the field normalization and disappears in the mutual information function $I(A,B)=S(A)+S(B)-S(AB)$ (we write $AB$ for the disjoint union of $A$ and $B$). Note the factor of $-i q$ attached to the conjugate fields, which will play a role later. At coincidence points there are additional singularities for the correlator which are not specified for the entropy. Alternatively, the identification of the exponential of the entropy can be made with the Euclidean correlator, without the $i$ factors. Thus, for the free massless fermion we have a kind of self-duality mediated by the entropy. In this particular example, the Renyi entropies are all proportional to the same quantity, $S_n((a_1,b_1)...(a_p,b_p))=(1+n)/(2n)\,\,S((a_1,b_1)...(a_p,b_p))$. As a consequence they can also be written in terms of correlators as in (\ref{fermi}).

The exponentials $e^{-(n-1)S_n(A)}=\textrm{tr} (\rho_A^n)$ of the Renyi entropies for integer index $n$ can also be written in a similar way for any QFT. In the Euclidean formulation of two dimensional QFT (and for a set $A$ lying in a single spatial line),  they are the vacuum expectation value of a product of twisting operators in a replicated model, located on the intervals end-points \cite{ccc,twist}.

If we extend this idea to other QFT, a more general formulation of the relation (\ref{fermi}) in two spacetime dimensions would take the form of a duality which from an entropy in the original theory gives place to the correlators of a new field,
\begin{equation}
e^{-\lambda S((a_1,b_1)...(a_p,b_p))}=\langle 0 \vert  \tilde{\Phi}(a_1) \Phi(b_1) ...\tilde{\Phi}(a_p)\Phi(b_p)  \vert 0 \rangle \,,\label{dua}
\end{equation}
where $a_1,b_1,a_2...b_p$ are spatially separated points in Minkowski space, ordered from left to right, but not necessarily on the same spatial line.  
As explained below, $\tilde{\Phi}(x)$ is the CPT conjugate of $\Phi(-x)$. The construction of the fields from the entropy (or the Renyi entropies) would follow uniquely if the Wightman axioms for real time correlators, or the Osterwalder Schrader axioms for the Euclidean ones are satisfied by the exponential of the entropy.  Here we focus on  one important ingredient in this construction which is the positivity relations for  the correlators. These are mapped to inequalities for the entropies. We prove these inequalities hold for the Renyi entropies of integer index in any dimension. We leave for a future work a more complete analysis on the validity of (\ref{dua}). This question is naturally related to another one, which is up to what extent the entropy determines the underlying QFT theory.

In more general terms we can ask whether is it possible to write the mutual information $I(A,B)$ in QFT in terms of a vacuum correlator of operators localized in $A$ and $B$. The mutual information is ultraviolet finite, and thus this question is meaningful in the renormalized theory.

The example of (\ref{fermi}) gives us the anzats 
 \begin{equation}
e^{\lambda I(A,B)}=\frac{
\langle 0 \vert {\cal O}_A\otimes {\cal O}_B \vert 0 \rangle}{\langle 0 \vert {\cal O}_A\vert 0 \rangle\langle 0 \vert {\cal O}_B \vert 0 \rangle}\,,\label{mutual}
\end{equation}
where $\lambda$ is some number and $\vert 0\rangle$ is the vacuum state.
This corresponds to 
\begin{equation}
e^{-\lambda S(A)}=\langle 0 \vert  {\cal O}_A \vert 0 \rangle\,,\label{ista}
\end{equation}
 and we have to take 
 \begin{equation}{\cal O}_{AB}={\cal O}_A\otimes {\cal O}_B\label{parts}
 \end{equation}
  for disjoint $A$ and $B$. Eq. (\ref{ista}) is very different from the defining relation $
 S(A)=-\langle 0 \vert  \log \rho_{A}\otimes 1_{-A} \vert 0 \rangle=-\textrm{tr} \rho_A \log (\rho_A)$, which is a consequence of $\rho_A=\textrm{tr}_{-A}\vert 0\rangle \langle 0 \vert $.

Note that the entropy $S(AB)$ for very distant regions $A$ and $B$ approaches the sum of the entropies of $A$ and $B$,  while the vacuum expectation value (vev) of the product of operators in $A$ and $B$ tends to the product of the vev´s. Thus, the exponential in (\ref{ista}) is exactly what is needed in order to respect the clustering properties of correlators and entropies.  

This mapping should also preserve Poincar\'e symmetry and causality. Lorentz invariance of $S(A)$ in a non chiral theory requires that the field $\Phi(x)$ in (\ref{dua}) is a scalar (it is the product of the two chiral components for the massless fermion in (\ref{fermi})). According to (\ref{parts}) ${\cal O}_A$ is a product of operators over the parts of $A$. Because of causality ${\cal O}_A$ should be the same for all spatial surfaces with the same boundary as $A$ \cite{ch2}. This suggests that the operator ${\cal O}_A$ is localized on the boundary of $A$. In more than one spatial dimension it may then be some kind of generalized Wilson loop, while in one spatial dimension one expects the products of  fields attached to the endpoints of the set intervals, as in (\ref{dua}).
 
\section{Real time reflection positivity}
An important requirement the entropy functions have to satisfy for the existence of the relations (\ref{mutual}-\ref{ista}), or more specifically (\ref{dua}), are the positivity properties of the correlators. These translate the Hilbert space positivity  of the scalar  product into the language of correlation functions.
Because the entanglement entropies are defined in Minkowski space, and in order to make contact with a more general quantum mechanical context, we introduce a real time reflection positivity property, in the real time formulation of QFT, rather than in the Euclidean one where reflection positivity is usually presented  \cite{refl}. Also, it is not known how to represent in the Euclidean framework the Renyi entropies of non-integer index, or the entropies for spatial sets which are not included in a single plane. 

 In a general quantum mechanical setting these inequalities can be derived within the Tomita-Takesaki modular theory (see for example \cite{haag}), which is as follows. Given a vector state\footnote{More precisely a cyclic and separating vector, that is, a vector such that ${\cal A} \vert 0 \rangle$ and ${\cal A}^\prime \vert 0 \rangle$ both span the whole Hilbert space, with ${\cal A}^\prime$ the commutant algebra of ${\cal A}$. This technical requirement holds for the applications of this paper.} $\vert 0 \rangle$ in an operator algebra ${\cal A}$, we can define the antilinear operator $S$ (not to be confused with the entropy) by
 \begin{equation}
 S {\cal O}\vert 0 \rangle={\cal O}^\dagger \vert 0 \rangle\,,
 \end{equation}
 for any ${\cal O}\in {\cal A}$. $S$ can be decomposed as $S=J \Delta^{\frac{1}{2}}$, with $J$ antiunitary and $\Delta$ self-adjoint and positive definite. 
  The crucial point is that $J$ maps the algebra ${\cal A}$ into its commutant algebra ${\cal A}^\prime$.
 One also has $\Delta \vert 0 \rangle=\vert 0 \rangle$, $J \vert 0 \rangle=\vert 0 \rangle$ and $J \Delta = \Delta^{-1} J$. Then it follows, writing $\bar{{\cal O}}=J{\cal O}J$ for the "reflected" operator,
 \begin{equation}
 \langle 0\vert {\cal O} \bar{{\cal O}} \vert 0 \rangle=\langle 0\vert {\cal O} J {\cal O} \vert 0 \rangle=\langle 0\vert {\cal O} \Delta^{\frac{1}{2}} S {\cal O} \vert 0 \rangle=\langle 0\vert {\cal O} \Delta^{\frac{1}{2}} {\cal O}^\dagger \vert 0 \rangle\geq 0\,,\label{rere}
 \end{equation}
 for any ${\cal O}\in {\cal A}$. 
 This is a general quantum mechanical reflection positivity property. 
 
 The connection with QFT is given by the fact discovered in \cite{aa}. Let us call the "wedge" to the set ${\cal W}$ of points $(t,x)$ in two spacetime dimensions with $x>0$, $\vert t\vert < x$. In relativistic QFT there is an algebra ${\cal A}_{\cal W}$ of operators localized in the wedge ${\cal W}$. For the vacuum state the modular reflection operator $J$ corresponding to the wedge is given in two spacetime dimensions by the CPT operator \cite{haag,aa}. The remarkable fact is then that the operator $J$ acts geometrically on the operator subalgebras as an inversion of coordinates $(t,x)\rightarrow (-t,-x)$ (see figure 1).
 Thus we have the positivity (reflection positivity) of 
\begin{equation}
\langle 0\vert  {\cal Q}(J{\cal Q}J)\vert 0\rangle\geq 0\,,\label{ref}
\end{equation}
for any operator ${\cal Q}$ localized inside ${\cal W}$.

 A similar situation holds in any dimension. In four dimensions the operator $J$ corresponding to the wedge (formed by the points $(t,x,y,z)$ with $x>0$, $\vert t\vert<x$) is given by the CPT operator followed by a rotation of angle $\pi$ around the $x$ axis. Thus, $J$ maps the operators at the point $(t,x,y,z)$ to operators at the reflected point $(-t,-x,y,z)$.  

\begin{figure}[t]
\centering
\leavevmode
\epsfysize=7cm
\epsfbox{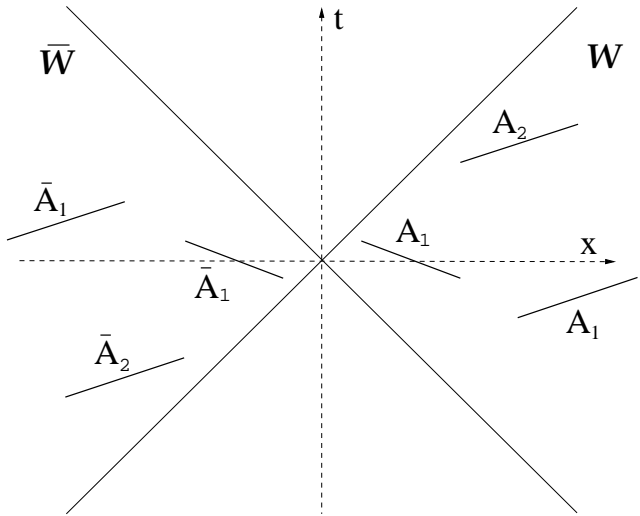}
\bigskip
\caption{Here the two-interval spatial set $A_1$ and the single spatial interval $A_2$ are included in the wedge ${\cal W}$ ($x>0$, $\vert t\vert < x $). Their reflected images are $\bar{A}_1$ and $\bar{A}_2$. The null lines $x=\pm t$ form the boundaries of the wedge ${\cal W}$ and the reflected wedge $\bar{{\cal W}}$. }
\end{figure}

 The operator $J$ takes the algebra of operators in a region $A$ to the one in the region $\bar{A}$, which is the reflected image of $A$. Since it keeps the vacuum invariant $J\vert 0 \rangle=\vert 0\rangle$, we have $S(A)=S(\bar{A})$, and according to (\ref{ista}) we can take ${\cal O}_{\bar{A}}=J {\cal O}_A J$. This explains the factors of $-iq$ in (\ref{fermi}), since we have for a chiral fermion field $J \Psi(\vec{x})J^{-1}=-i q \Psi^\dagger(-\vec{x})$. 
 
 Note that in (\ref{fermi}) the field $\Psi(x)$ with $x\in {\cal W}$ does not commute with the fermion operators in $(-{\cal W})$, but rather anticommutes with them. Thus, it does not belong to the algebra ${\cal A}_{\cal W}$ generated by the observables localized in ${\cal W}$, and an extension of the modular reflection $J$ is needed (this is just given by the CPT operator acting on fermions, but one has to use $J {\cal Q} J^{-1}$ instead of $J {\cal Q} J$, since $J\neq J^{-1}$ for fermios). What is important to the argument is that the operator ${\cal O}_A$ for a region $A$ in ${\cal W}$ must be neutral and belong to ${\cal A}_{\cal W}$, whether the individual operators attached to the boundaries are observables or not. In more dimensions, this is also the case of Wilson loop type operators.

 Then we insert in the real time reflection positivity relation (\ref{ref}) the linear combination ${\cal Q}=\sum_{i=1}^{m+1} \alpha_i {\cal O}_{A_i}$ with arbitrary coefficients $\alpha_i$, and where the sets $A_i\subseteq {\cal W}$ can have any number of components, and are all in the same wedge ${\cal W}$.  We learn from (\ref{ref}) that the matrix of $e^{-\lambda S(A_i \bar{A}_j)}$ should be positive definite, that is 
\begin{equation}
\det\left(\{e^{- \lambda S( A_i\bar{A}_j)}\}_{i,j=1,..,m+1}\right)\geq 0\,,\label{det}
\end{equation}
or equivalently 
\begin{equation}
\det\left(\{e^{ \lambda I( A_i,\bar{A}_j)}\}_{i,j=1,..,m+1}\right)\geq 0\,,
\end{equation}
for every integer $m\geq 1$ and collection of sets $A_i\subseteq {\cal W}$. These sets need not be spacelike separated to each other (see figure 1). 

The case of two sets $A$ and $B$ ($m=1$) gives the linear inequality
\begin{equation}
 S(A\bar{B})+S(B\bar{A})=2 S(A\bar{B})\geq S(A\bar{A})+S(B\bar{B})\,.\label{dos}
 \end{equation}
 This is generally independent of strong subadditivity ($
S(XZ)+S(YZ)\ge S(Z)+S(XYZ)$
 where $X$, $Y$, and $Z$ are non intersecting sets) 
 though it is a particular case of it when $A\subseteq B$ or $B\subseteq A$. For the entropy, compatibility with strong subadditivity requires $\lambda\geq 0$. We show below that these inequalities are valid in any dimension for the Renyi entropies of integer index $n$, taking  $\lambda=(n-1)$.

\section{Reflection positivity inequalities for the Renyi entropy in quantum mechanics}

The formulas (\ref{det}) can be interpreted in more general terms. Let $\rho$ be an invertible density matrix in a general quantum mechanical system of Hilbert space ${\cal H}_1$. Suppose ${\cal H}_1$ contains subsystems $A_1$,...$A_{m+1}$, which does not necessarily have commuting algebras. Let $\vert 0 \rangle$ be a purification of $\rho$ in the space ${\cal H}_1\otimes{\cal H}_2$, where ${\cal H}_2$ is a copy of ${\cal H}_1$. That is, $\rho=\textrm{tr}_{{\cal H}_2}\vert 0\rangle\langle 0\vert$. 
 Then consider the state $\vert 0\rangle$ and the algebra of all the operators acting on the first factor of ${\cal H}_1\otimes{\cal H}_2$. The corresponding modular reflection $J$ maps the original subsystems $A_1,...,A_{m+1}$ to subsystems of the second tensor product factor, $\bar{A}_1,...,\bar{A}_{m+1}$. The relations  (\ref{det}) make then sense in this more general context, where $S(A_i \bar{A}_j)$ is the entropy in the joint subsystem $A_i \bar{A}_j$.    
 
Let us describe in more detail the elements involved in (\ref{det}) for a general finite dimensional system. We can write the state $\vert 0\rangle$ as a Schmidt decomposition in ${\cal H}_1\otimes {\cal H}_2$,
\begin{equation}
\vert 0 \rangle = \sum_p \sqrt{\lambda_p}  \vert p \,\tilde{p}\rangle\,.\label{laba}
\end{equation}
 The $\lambda_p$, $p=1,...,d$ are the eigenvalues of $\rho=\textrm{tr}_{{\cal H}_2}\vert 0\rangle \langle 0 \vert$, which is the reduced density matrix to the ${\cal H}_1$ factor. The vectors $\vert p\rangle$ are the corresponding eigenvectors. However, the orthonormal base $\{\vert\tilde{p}\rangle\}$ for ${\cal H}_2$ in (\ref{laba}) is arbitrary, and different basis correspond to different purifications of $\vert 0 \rangle$.   
 
It is not difficult to check that the modular operators $\Delta$ and $J$ corresponding to the state $\vert 0\rangle$ in the algebra of the matrices acting on ${\cal H}_1$ are given by
\begin{eqnarray}
\Delta&=&\sum_{p,q} \frac{\lambda_p}{\lambda_q}\vert p \,\tilde{q}\rangle\langle p\, \tilde{q}\vert\,,\\
J&=&\sum_{p q} \vert p \,\tilde{q}\rangle \langle q\,\tilde{p}\vert  *\,.\label{jota}
\end{eqnarray}
Then the reflection $J$ is a product of a transposition of the basis in ${\cal H}_1$ and ${\cal H}_2$, with the operator $*$ of complex conjugation of the vector components written in the basis $\{\vert p \tilde{q} \rangle\}$.  

Given a subsystem $A$ of ${\cal H}_1$, let us have a decomposition  ${\cal H}_1={\cal H}_{A}\otimes {\cal H}_{B}$. Let the corresponding decomposition of the basis vectors be
\begin{equation}
\vert p \rangle=\sum_{k,l} \beta^{(A)p}_{k l}\vert k l\rangle\,,\label{quant}
\end{equation}
where $\beta^{(A)}$ is a unitary transformation,
\begin{equation}
\sum_{k,l}\beta^{(A)p}_{k l}\left(\beta^{(A)q}_{k l}\right)^*=\delta_{p,q} \,,\hspace{2cm} \sum_{p}\beta^{(A)p}_{k l}\left(\beta^{(A)p}_{k^\prime l^\prime}\right)^*=\delta_{k,k^\prime} \delta_{l,l^\prime}\,.
\end{equation}
It is convenient to take a decomposition of ${\cal H}_2$ given by
\begin{equation}
\vert \tilde{p} \rangle=\sum_{k,l} \left(\beta^{(A)p}_{k l}\right)^*\vert \underline{k l}\rangle\,.
\end{equation}
This relation just defines the orthonormal basis $\vert \underline{k l}\rangle $. Note that this is not exactly homologous to (\ref{quant}). 
In this basis the conjugation $J$ (eq. (\ref{jota})) acts as the complex conjugation $\hat{*}$ of components (in this new basis) followed by transposition, 
\begin{equation}
J=  \sum_{k,l,k^\prime,l^\prime} \vert k l \underline{ k^\prime  l^\prime }\rangle \langle   k^\prime l^\prime \underline{k l}\vert  \,\,\,\,  \hat{*}\,.
\end{equation}
This allows us to identify easily the reflected subsystem $\bar{A}$ in ${\cal H}_2$ as the factor spanned by the vectors $\vert \underline{k}\rangle$ in the basis $\vert \underline{k l}\rangle $.

Now, suppose we have a collection of subsystems $A_1,...,A_{m+1}$ in ${\cal H}_1$. We can write $\vert 0\rangle$ in a mixed basis, formed by the decomposition ${\cal H}_1={\cal H}_{A_i}\otimes{\cal H}_{B_i}$ of the first factor, and the decomposition ${\cal H}_2={\cal H}_{\bar{A}_j}\otimes{\cal H}_{\bar{B}_j}$ of the second one. We have
\begin{equation}
\vert 0\rangle= \sum_{p, k_i, l_i,
k_j,l_j}\sqrt{\lambda_p}\,\beta^{(A_i)p}_{k_i l_i} \left(\beta^{(A_j)p}_{k_j l_j}\right)^* \vert k_i l_i \underline{ k_j l_j }\rangle \,.
\end{equation}
The partial traces of $\vert 0 \rangle \langle 0 \vert$ give place to the reduced density matrices. We have,
\begin{equation}
\rho_{A_i \bar{A}_j}=\sum_{p,q} {\cal O}_{A_i}^{pq} \otimes\bar{{\cal O}}_{A_j}^{pq} \,,\label{dm}
\end{equation}
where 
\begin{equation}
{\cal O}_{A_i}^{pq}=(\lambda_p \lambda_q)^{\frac{1}{4}} 
\sum_{
k_i, k_i^\prime,l_i}\beta^{(A_i)p}_{k_i l_i} \left(\beta^{(A_i)q}_{k_i^\prime l_i}\right)^* \vert k_i \rangle \langle k_i^\prime \vert\,,
\end{equation}
and
\begin{equation}
\bar{{\cal O}}^{pq}_{A_j}=J {\cal O}^{pq}_{A_j} J=(\lambda_p \lambda_q)^{\frac{1}{4}} 
\sum_{
k_j, k_j^\prime,l_j}\left(\beta^{(A_j)p}_{k_j l_j}\right)^* \beta^{(A_j)q}_{k_j^\prime l_j} \vert\underline{ k_j} \rangle \langle \underline{k_j^\prime} \vert\,.
\end{equation}
At this point it is evident that all the different purifications give place to the same entropies $S(A_i\bar{A}_j)$, which are in fact functions of the density matrix $\rho$ and the subsystems $A_i$ and $A_j$ in ${\cal H}_1$.

We can exploit the particular structure (\ref{dm}) for the density matrices.  We apply the reflection positivity (\ref{rere}) to the general linear combination 
\begin{equation}
{\cal Q}^{p_1,q_1,...,p_{n-1}q_{n-1}}=\sum_k \alpha_k {\cal O}^{p_1q_1}_{A_k}...{\cal O}^{p_{n-1}\,q_{n-1}}_{A_k}\,,
\end{equation} 
with arbitrary coefficients $\alpha_k$, and then sum over $p_1,...,p_{n-1},q_1,...q_{n-1}$. We get 
\begin{equation}
\det \left(  \left\{ \textrm{tr}\rho_{A_i \bar{A}_j}^{n}\right\}_{i,j=1,...,m+1}   \right)=\det \left(\left\{e^{-(n-1) S_{n}(A_i,\bar{A}_j)}\right\}_{i,j=1,...,m+1}\right)\geq 0\,,\hspace{.2cm} n=1,2,.. \,.\label{trece}
\end{equation}
The case $m=1$ gives place to the linear inequality 
\begin{equation}
2 S_n(A\bar{B})\ge S_n(A\bar{A})+S_n(B\bar{B})\,.
\end{equation}   
Eq.(\ref{trece}) is eq. (\ref{det}) for the Renyi entropies $S_n$ of integer index $n$, and for a specific coefficient $\lambda=(n-1)$. It gives an infinite series of inequalities for these Renyi entropies which are valid in a general quantum system.  This gives support to the representation of these entropies in QFT as correlators of twisting operators in Euclidean space \cite{twist}. 

It is known (Schur product theorem) that if $\{M_{ij}\}$ is positive definite, the matrices $\{(M_{ij})^k\}$, with entries which are integer powers of the ones of $M$ is also positive definite. From (\ref{trece}) it follows that  $\det (\{ (\textrm{tr}\rho_{A_i\bar{A}_j}^n)^s\}_{i,j=1,...,m+1})\geq 0$ for all $n$ and $s$ positive integer. We would like to know the conditions allowing to extend these inequalities for other real $s\geq 0$ (equivalently to $\lambda\neq (n-1)$) and other values of $n\geq 1$.  In a general quantum system even if (\ref{trece}) holds, the extension to the entropy case ($n\rightarrow 1$), or the infinite divisible case (see below) $s\rightarrow 0$, may fail. We have found some counterexamples to both using randomly generated matrices in low dimensions. However, the volume of the space of density matrices violating the inequalities in those cases seems to be rather small.  

\section{Kall\'en-Lehmann representation for the single interval case}

For a single interval the full set of inequalities can be solved giving place to a Kall\'en-Lehmann type representation for the two point function. From the assumption that $\exp(-\lambda S(a,b))$ is a correlator, we have 
\begin{equation}
e^{-\lambda S(a,b)}=\int d^2\,p\,\, g(p^2)\theta(p^0) e^{-i p.(a-b)}\,,
\end{equation} 
with $\theta(p^0)g(p^2)$ Lorentz invariant and with support in the positive light cone $g(p^2)=\theta(p^2) g(p^2)$. This is due to the positive energy  condition for the intermediate physical states, 
\begin{equation}
\langle 0\vert\tilde{\Phi}(a)\Phi(b) \vert 0 \rangle=\sum_p \langle 0\vert\tilde{\Phi}(a)\vert p\rangle \langle p\vert\Phi(b) \vert 0 \rangle=\sum_p \vert \langle p\vert\Phi(0) \vert 0 \rangle\vert^2
e^{i p (b-a)}\,,\label{sse}
\end{equation}
where we have used $\tilde{\Phi}(x)$ is the CPT conjugate of the scalar field $\Phi(x)$. Eq. (\ref{sse}) also shows the positivity of the spectral function $g(p^2)$, directly from the positivity of the Hilbert space metric.

 Changing variables $(p^0,p^1)=p(\cosh(\alpha),\sinh(\alpha))$, and integrating over the  boosts variable $\alpha$ we have the representation
\begin{equation}
e^{-\lambda S(a,b)}=\int_0^\infty dp \,p \int_{-\infty}^\infty d\alpha g(p^2) e^{i p \vert a-b\vert \sinh(\alpha)}=\int_0^\infty d p^2\,\, g(p^2) \,K_0(p \vert a-b\vert)\,,\label{repa}
\end{equation} 
with $K_0(x)$ the  modified Bessel function. This is the Kall\'en-Lehmann representation. 

We now show that the infinite many conditions (\ref{det}), given by Minkowskian reflection positivity on this function, are equivalent to $g(p^2)\ge 0$. The inequalities can be summarized as
\begin{eqnarray}
&&\int d^2a \,d^2b\,  d^2p\, \theta(p^0) g(p^2) e^{-i p.(a-b)} f(-a)f^*(b)=\int d^2a\, d^2b \, d^2p \,\theta(p^0) g(p^2) e^{i p.(a+b)} f(a)f^*(b)\nonumber \\
&&=\int d^2a d^2b \int dp^2 \, g(p^2) K_0(p \vert a+b\vert)f(a)f^*(b)\ge 0\,,\label{rep}
\end{eqnarray} 
where $f(x)$ is an arbitrary test function with compact support included in ${\cal W}$.
Now, using $K_0(x)=(1/2) \int d\alpha  e^{-x \cosh(\alpha)}$ we have, using Lorentz invariance of $g(p^2)$, 
\begin{eqnarray}
&&\int d^2a d^2b \int dp^2 \, g(p^2) f(a)f^*(b)\int d\alpha \,\,\frac{1}{2}\, e^{-p \vert a+b \vert \cosh(\alpha)}=\nonumber\\
&&\int d^2a d^2b \int d^2p \, \theta(p^1)g(-p^2) f(a)f^*(b)  e^{p ( a+b)}=\int d^2p \, \theta(p^1)g(-p^2) \check{f}(p)\check{f}^*(p) \ge 0\,,
\end{eqnarray}
where $\check{f}(p)=\int d^2a f(a)e^{p a}$, and the integrals over momentum in both sides of the last equation are over all $p\in {\cal W}$, with $p^2<0$. 
This shows the inequalities (\ref{rep}) reduce to $g(p^2)\ge 0$. 

It is interesting to note that even if (\ref{repa}) with positive $g(p^2)$ includes all the information from reflection positivity for the single interval, it still allows for features which should not be present for an entropy of the vacuum state. It allows for example for a linearly increasing entropy at large distances (just take a $g(p^2)$ different from zero for $p^2\ge \Lambda$, with $\Lambda$ some gap). A related issue is that the reflection positivity inequalities imply $S^\prime(x)\ge 0$ and $S^{\prime\prime}(x)\le 0$, which also follow from strong subadditivity for space like intervals, but do not imply the stronger relation (entropic c-theorem \cite{nu}) $x S^{\prime\prime}(x)+S^\prime(x)\le 0$, which is a consequence of strong subadditivity for non collinear intervals. This last relation forbids linearly increasing entropies. Since the strong subadditivity is not a property (in general) for the Renyi entropies with $n\neq 1$ we still do not know what enforces non linear increasing Renyi entropies.         

For a massive theory one should have a contribution to $g(p^2)$ proportional to a delta function $\delta(p^2)$, in such a way to allow for saturation of the entropy at large distances. The remaining non-zero part of $g(p^2)$ should be located at $p^2\ge 4 M^2$, where $M$ is the physical mass of the theory. This gives a characteristic exponential decay  $\sim e^{-2 Mx}$ for the subleading terms on the entropies when approaching saturation \cite{fermion,twist,cc2}. 
The short distance behavior of the entropies is governed by the large $p$ behavior of $g(p^2)$. The general short distance behavior for one interval is $S_n(x)\sim C\frac{ (n+1)}{6 n}\log(x)$ with $C$ the Virasoro central charge of the conformal ultraviolet fixed point \cite{ccc}.  
Writting $g(p^2)=\textrm{cons}\,\, p^\gamma$ for large $p$, we have for the Renyi entropies $\gamma+2=\lambda  \frac{n+1}{6 n}\, C$, .

\section{Infinite divisibility}
From Schur's theorem, if (\ref{det}) is satisfied for $\lambda$ it is also satisfied for $k \lambda$ with $k$ positive integer. If $\lambda$ can be taken as small as we want, the positive definite matrix in (\ref{det}) is called infinitely divisible (see for example \cite{bha}), and the inequalities hold automatically for any $\lambda > 0$. In this case, expanding in series for $\lambda\rightarrow 0$ the  inequalities (\ref{det}) simplify to  
\begin{eqnarray}
\det B\geq 0\,, \hspace{.8cm} B_{ij}=S(A_i\bar{A}_{j+1})+S(A_{i+1}\bar{A}_j)-S(A_i\bar{A}_j)-S(A_{i+1}\bar{A}_{j+1})\nonumber\\
=I(A_i,\bar{A}_j)+I(A_{i+1},\bar{A}_{j+1})-I(A_i,\bar{A}_{j+1})-I(A_{i+1},\bar{A}_j)
\,,\hspace{.4cm} i,j=1,...,m\,.
\label{uh}\end{eqnarray}
 These are still quite complicated in general. The first non-linear inequality reads explicitly for $A$, $B$ and $C$ in ${\cal W}$
 \begin{eqnarray}
&&2 S(A\bar{B})S(A\bar{C})+2 S(A\bar{B})S(B\bar{C})+2 S(B\bar{C})S(A\bar{C})+ S(A\bar{A})S(B\bar{B})+S(A\bar{A})S(C\bar{C})\nonumber \\\nonumber &&
+S(B\bar{B})S(C\bar{C})\nonumber 
\geq S(A\bar{B})^2+S(A\bar{C})^2+S(B\bar{C})^2 +2 S(A\bar{B})S(C\bar{C}) +2 S(A\bar{C})S(B\bar{B})\\ 
&&
+2 S(B\bar{C})S(A\bar{A})\,. 
\end{eqnarray}
Here we have used $S(A_i \bar{A}_j)=S(A_j \bar{A}_i)$.

Thus, we have a polynomial inequality of degree $m$, involving $m+1$ different subsystems and their reflected counterparts, for each positive integer $m$. Note that if the inequalities (\ref{trece}) can be extended down to $n=1$ in a QFT then the exponentials of the entropy are automatically infinite divisible. 

Infinite divisibility means in this context that we can take any $\lambda> 0$ and there is no difference with respect to reflection positivity. In other words, the powers of a correlator are again correlators. The interpretation in (\ref{dua}) is however greatly dependent on $\lambda$, since for example, the field scaling dimension at the conformal point is proportional to this parameter.

For the entropy function of the free massless fermion (\ref{fermi}) we have infinite divisibility for the matrices (\ref{det}) for any $\lambda$ and the inequalities (\ref{uh}) hold. This follows from reflection positivity, since we can write for any $\lambda >0$
\begin{equation}
 \tilde{c}^{p} e^{- \lambda S((a_1,b_1)...(a_p,b_p))}=\langle 0 \vert  :e^{i\sqrt{\frac{2\pi \lambda}{3}} \phi(a_1)}:  :e^{-i\sqrt{\frac{2\pi \lambda}{3}} \phi(b_1)}: ... :e^{i \sqrt{\frac{2\pi \lambda}{3}}\phi(a_p)}: :e^{-i \sqrt{\frac{2\pi \lambda}{3}}\phi(b_p)}:  \vert 0 \rangle \,,\label{fem}
\end{equation}
in terms of vertex (exponential) operators constructed with a free massless scalar field $\phi(x)$ \cite{fermion}. The right hand side gives the left hand one since we have 
\begin{equation}
\langle 0 \vert e^{i \int dx f(x) \phi(x)}\vert 0 \rangle=e^{\frac{1}{8 \pi}\int dx dy f(x) \log \vert x-y \vert f(y)}\,.\label{quad}
\end{equation}

Further examples of analytical results for the Renyi entropies in two dimensional theories include the single interval entropies for free massive scalar and Dirac fields. For integer $n$ these are given as a finite sum of terms involving the solutions of Painlev\'e non-linear differential equations \cite{fermion,pain}. The Renyi entropies for real $n$ and the entropy ($n=1$) case are given in terms of integrals involving Painlev\'e functions \cite{analitic}. We have tested numerically the infinite divisibility inequalities in the integer $n$ case. Up to what we have checked the inequalities (\ref{uh}) are obeyed in this case. It is remarkable that for the free massive fermion we still have a formula analogous to (\ref{fem}), giving the integer $n$ Renyi entropies in terms of a product of correlators of vertex operators (for details see \cite{fermion}),
\begin{equation}
 \tilde{c}^{p} e^{- (n-1) S_n((a_1,b_1)...(a_p,b_p))}=\prod_{k=-\frac{n-1}{2}}^{\frac{(n-1)}{2}}\langle 0 \vert  :e^{i\sqrt{4\pi}\frac{k}{n} \phi(a_1)}:   ...  :e^{-i \sqrt{4\pi}\frac{k}{n}\phi(b_p)}:  \vert 0 \rangle \,,
\end{equation} 
but in this case the scalar field $\phi(x)$ is no longer free, but belongs to the sine-Gordon theory at the free fermion point.   Thus, it is no longer possible to use (\ref{quad}) to obtain a representation of the powers of the correlators in terms of correlators. If confirmed, infinite divisibility would then be an intriguing  property of the Painlev\'e functions, perhaps related to a different type of operators.

The exact Renyi entropies of integer index $n$ for two intervals are also known for other models.  The Renyi entropy $S_2$ for the case of a compactified massless scalar was obtained in \cite{nue}. In \cite{tema} this result was extended and the formula for $S_n$ for two intervals and $n$ integer greater than two was found. The critical Ising model is studied in \cite{nue1} where $S_2$ is given.  
In general, conformal invariance implies the entropies of two intervals can be written as \cite{nu,nue}
\begin{equation}
e^{-(n-1)S_n}=k^2 \left(x (a_2-b_1)(b_2-a_1)\right)^{-\frac{C}{6}(n-\frac{1}{n})}F_n(x)\,,
\end{equation}
where $F_n(x)=F_n(1-x)$, $F(0)=1$, is a function of the cross ratio $x=\frac{(b_1-a_1)(b_2-a_2)}{(a_2-a_1)(b_2-b_1)}$, $k$ is a constant, and $C$ is the Virasoro central charge. Writing $q=\frac{C}{6}(n-\frac{1}{n})$, the linear inequality gives for the function $F$ 
\begin{eqnarray}
\frac{\partial\left(\frac{F_n(x)}{(1-x)^q}\right)}{\partial x}\ge 0\,,\\
\frac{F_n(x)}{(1-x)^q} \frac{F_n(y)}{(1-y)^q}\ge \left(\frac{F_n(z)}{(1-z)^q}\right)^2\,,
\end{eqnarray}
where this last inequality holds for any $x$ and $y$, with $z=\frac{2 \sqrt{x y}}{1+\sqrt{1-x}\sqrt{1-y}+\sqrt{x y}}\in (x,y)$. 

In general the functions $F_n(x)$ for the known examples are given by expressions involving theta functions and their inverses. For the case of $S_2$ for the compactified scalar with parameter $\eta=1/2$ \cite{nue}, and for the Ising model given in \cite{nue1}, $F_2(x)$ simplifies and is given in terms of algebraic functions. For these cases we have tested numerically the linear inequalities and some of the non linear ones (\ref{trece}) using randomly generated sets. As expected the inequalities hold, what is consistent with $\textrm{tr}\rho^n$ being given by a vacuum expectation value of a twisting operator \cite{twist}. However, infinite divisibility does not hold. Thus, it seems infinite divisibility is not a general property and it is not expected to hold in all cases. The examples may indicate it would be related to the theory being free. A small caveat remains though, since it is not completely clear in which sense the Ising model entropies (and the Renyi entropies calculated in \cite{nue}) can be interpreted as entropies for regions in a QFT, with a tensor product structure of the local Hilbert spaces. A better understanding of this point would be desirable. 

\section{Final remarks}

The search for a basis to decide whether the entanglement entropy of the vacuum could be written in terms of an expectation value have led us to use a positivity relation, a real time reflection positivity, which is dictated by the Tomita Takesaki theory, and which is different from both, the Euclidean reflection positivity and the usual Hilbert space positivity of correlators in the Minkowskian framework. This seems to suggest a different axiomatic formulation of QFT in a mixed scheme somewhat intermediate between the Wightman and the Osterwalder-Schrader schemes. This would have in common with the Wightman scheme the use of Minkowskian correlators, and Lorentz symmetry, but the correlators are restricted to spatial separations of the points, the so called Jost points. This is appropriate for the entropies, which are defined in Minkowski space and are Lorentz invariant, but are restricted to spatial sets. Instead of the spectrum condition and Wightman positivity we should have analyticity and Minkowskian reflection positivity, somewhat in the fashion of the Euclidean axioms. It would be interesting to explore this possibility.    
  
In a different direction, the existence of new geometric non-linear inequalities for the entropy in QFT may be useful in trying to prove the c-theorem in more than two dimensions. The strong subadditive inequality leads to a c-theorem for the entanglement entropy in two spacetime dimensions, but its implementation in more dimensions is impeded precisely by its linear character \cite{ct}. 

There are also interesting scenarios where the predictions for the Renyi entropies  could be tested with the inequalities of this paper. For example, this is the case of the results of a recent paper \cite{he} where some Renyi entropies of two dimensional conformal theories are obtained by holographic methods \cite{rt}. 

Finally we note that the study of the strong subadditive property of the entanglement entropy in the context of the Maldacena duality has lead to the conjecture that a similar property might also hold for some Wilson loop operators \cite{ta}. The inequalities for the entropy discussed in this paper have also a natural geometric form for the Wilson loops, just expressing the reflection positivity of the Wilson loop operators  (the linear inequality is studied in \cite{bachas}). In particular, the strong subadditive property holds for a pair of loops which are reflected to each other with respect to a plane. 

\section*{Acknowledgments} 
I would like to thank specially Ernesto Huerta for his constant and warm encouragement, and Marina Huerta for discussions.
It is also a pleasure to thank F. Verstraete and the Erwin Schrodinger Institute for the kind invitation to participate in the workshop "Quantum Computation and Quantum Spin Systems", and the program "Entanglement and Correlations in Many body Quantum Mechanics", where this investigation began.  
 The author has benefited from correspondence with Pasquale Calabrese who suggested testing the inequalities on the results of \cite{nue,tema,nue1}. 
This work was partially supported by CONICET and Universidad Nacional de Cuyo, Argentina.

\end{document}